\documentclass[runningheads]{llncs}
\usepackage{graphicx}
\usepackage{amsmath}
\usepackage{multirow}
\usepackage{makecell}
\usepackage{amssymb}
\usepackage{booktabs}
\usepackage{cite}
\usepackage{xcolor}

\begin{document}
\titlerunning{Multimodal Information Fusion for Glaucoma and DR Classification}

\title{Multimodal Information Fusion for Glaucoma and Diabetic Retinopathy Classification}
%
%


\author{Yihao Li\inst{1,2}
\and
Mostafa El Habib Daho \inst{1,2}
\and
Pierre-Henri Conze\inst{1,3} 
\and
Hassan Al Hajj\inst{1,2}
\and
Sophie Bonnin\inst{4} 
\and
Hugang Ren\inst{5} 
\and
Niranchana Manivannan\inst{5} 
\and
Stephanie Magazzeni\inst{5} 
\and
Ramin Tadayoni\inst{6} 
\and
Béatrice Cochener\inst{1,2,7} 
\and
Mathieu Lamard\inst{1,2} 
\and
Gwenolé Quellec\inst{1} 
}

\authorrunning{Y. Li et al.}
%
\institute{
LaTIM UMR 1101, Inserm, Brest, France \and
University of Western Brittany, Brest, France \and
IMT Atlantique, Brest, France \and
Ophthalmology Department, Rothschild Foundation Hospital, Paris, France
\and
Carl Zeiss Meditec Inc, Dublin, CA, United States
\and
Ophthalmology Department, Lariboisière Hospital, AP-HP, Paris, France
\and
Ophthalmology Department, CHRU Brest, Brest, France
}
\maketitle              

\begin{abstract}

Multimodal information is frequently available in medical tasks. By combining information from multiple sources, clinicians are able to make more accurate judgments. In recent years, multiple imaging techniques have been used in clinical practice for retinal analysis: 2D fundus photographs, 3D optical coherence tomography (OCT) and 3D OCT angiography, etc. Our paper investigates three multimodal information fusion strategies based on deep learning to solve retinal analysis tasks: early fusion, intermediate fusion, and hierarchical fusion. The commonly used early and intermediate fusion are simple but do not fully exploit the complementary information between modalities. We developed a hierarchical fusion approach that focuses on combining features across multiple dimensions of the network, as well as exploring the correlation between modalities. These approaches were applied to glaucoma and diabetic retinopathy classification, using the public GAMMA dataset (fundus photographs and OCT) and a private dataset of PLEX®Elite 9000 (Carl Zeis Meditec Inc.) OCT angiography acquisitions, respectively. Our hierarchical fusion method performed the best in both cases and paved the way for better clinical diagnosis.\\
Multimodal information is frequently available in medical tasks. By combining information from multiple sources, clinicians are able to make more accurate judgments. In recent years, multiple imaging techniques have been used in clinical practice for retinal analysis: 2D fundus photographs, 3D optical coherence tomography (OCT) and 3D OCT angiography, etc. Our paper investigates three multimodal information fusion strategies based on deep learning to solve retinal analysis tasks: early fusion, intermediate fusion, and hierarchical fusion. The commonly used early and intermediate fusions are simple but do not fully exploit the complementary information between modalities. We developed a hierarchical fusion approach that focuses on combining features across multiple dimensions of the network, as well as exploring the correlation between modalities. These approaches were applied to glaucoma and diabetic retinopathy classification, using the public GAMMA dataset (fundus photographs and OCT) and a private dataset of PLEX\circledR Elite 9000 (Carl Zeis Meditec Inc.) OCT angiography acquisitions, respectively. Our hierarchical fusion method performed the best in both cases and paved the way for better clinical diagnosis.

\end{abstract}
\section{Introduction}
Glaucoma and diabetic retinopathy (DR) are two of the leading causes of blindness and visual impairment in the world. The glaucomatous neurodegeneration causes a disconnection between the retina and the brain, resulting in irreversible blindness. By 2040, around 111.8 million people are expected to suffer from glaucoma \cite{THAM20142081}. The DR mutilates the retinal blood vessels of diabetic patients. Diabetic retinopathy consists of two major types: non-proliferative diabetic retinopathy (NPDR) and proliferative diabetic retinopathy (PDR) \cite{8869883}. By 2030, there will be 454 million DR patients worldwide \cite{SAEEDI2019107843}.

In recent years, algorithms for diagnosing glaucoma and DR have emerged with the development of deep learning and improved computer equipments. Fundus photography and optical coherence tomography (OCT) are the two most cost-effective screening tools for glaucoma and DR \cite{https://doi.org/10.48550/arxiv.2202.06511}. For two-dimensional fundus photographs, powerful convolutional neural networks (CNN) such as ResNet or GoogleNet Inception models, were used to achieve pathology detection \cite{Shibata2018,10.1371/journal.pone.0207982,10.1167/tvst.8.6.4}. It should be noted that 2D fundus data are more accessible than other modalities, so data-sets are generally larger, and thus, models can be trained more efficiently. OCT data are more sensitive to structural pathological features. Both 3D-CNN networks and 2D-CNN networks operating on 2D slices, were used to achieve feature extraction from OCT volumes \cite{ASAOKA2019136,Muhammad2017,PERDOMO2019181}. In addition, optical coherence tomography angiography (OCTA) is a new, non-invasive imaging technique that generates volumetric angiography images in seconds. It can display both structural and blood flow information \cite{deCarlo2015}. The effectiveness of CNN networks in classifying DR using OCTA data was also demonstrated \cite{Ryu2021}.

All the previous algorithms are usually based on information from only one modality. However, multi-modality screening is often recommended to reach a more accurate and reliable diagnosis \cite{https://doi.org/10.48550/arxiv.2202.06511}. This is why multimodal algorithms are needed in ophthalmic pathology diagnosis.

This paper presents three fusion algorithms for multimodal data in ophthalmology: early, intermediate and hierarchical fusion. They enable the fusion of 2D and 3D modal data. Specifically, the innovative hierarchical fusion algorithm we developed (Fig.~\ref{fig1}) achieves excellent glaucoma and DR classification results.

\section{Methods}
This section will explore three approaches to multimodal fusion in ophthalmology: early fusion, intermediate fusion, and hierarchical fusion. We will examine the challenges of applying different fusion methods to ophthalmic data and the structural aspects of our network.

\subsection{Early fusion}
In early fusion, also called input-level fusion, data from different modalities are fed into a classification network as different channels \cite{ZHOU2019100004}. Specifically, multi-modality images are fused channel by channel to form multi-channel inputs. Then, a classification network is trained to learn a fused feature representation from these inputs. Many of today's medical fusion strategies 
use early fusion \cite{CLERIGUES2020105521,10.1007/978-3-319-75238-9_25}.

Let $X \times Y \times Z$ denote the size of the 3D volumes in voxels. In the early-fusion solution, the 2D images are resized to $X \times Y$ pixels and duplicated $Z$ times, to form a $X \times Y \times Z$ voxel channel. Feature extraction from the multimodal input of size $C \times X \times Y \times Z$, where $C$ denotes the number of channels, is then performed using a 3D-CNN network. In addition, the alignment of different modalities is crucial to early fusion.

Early fusion is a simple method, but it is not very effective due to the semantic gap between the modalities of ophthalmic data.
For example, fundus photographs give an overall en-face view of the retina, in 2D and OCT volumes provide structural information about the retina in 3D. However, there is a significant gap between these two modalities regarding the equipment used to capture them, imaging methods, and data information.
In particular, when we convert 2D data into 3D volumes, we cannot guarantee that the modalities are accurately aligned.

\subsection{Intermediate fusion}

In contrast with early fusion, intermediate fusion does not assume spatially aligned modalities. Instead, each modality data is used as an input to a single classification branch, and the outputs from each branch are integrated to produce a final result \cite{GAO201749, 10.1371/journal.pone.0237674}. The intermediate fusion strategy fuses features before the final decision layer. In contrast, late fusion fuses the decision results, ignoring any correlation between the different modalities \cite{doi:10.1080/0952813X.2019.1653383}.

As we use different independent branches to extract feature information from each modality, we do not need to consider the consistency of the input data. Using different 2D and 3D CNN branches to extract different features for 2D and 3D data is possible. 

Intermediate fusion is a simple yet effective method for feature fusion. The method effectively bridges the significant gaps between different modalities in ophthalmology (2D fundus images and 3D OCT or OCTA volumes). In particular, most participants in Task1 of the GAMMA Challenge employed this method to classify glaucoma and achieved good results \cite{https://doi.org/10.48550/arxiv.2202.06511}. Nevertheless, as intermediate fusion is a mere concatenation of high-dimensional features, the correlation information inevitably gets lost, adversely impacting classification performance.

\subsection{Hierarchical fusion}

In this work, we have extended the network structure of intermediate fusion to address its shortcomings.
Like intermediate fusion, hierarchical fusion works by using each modality image as an input of a single classification branch, then fusing these learned individual feature representations in the deeper layers of the network. However, unlike intermediate fusion, an additional branch performs feature fusion at different scales. A decision layer is then applied to the fused result to obtain the final label \cite{ZHANG2020108795}.

Fusion between modality-specific features of different dimensions in a network structure is challenging. Prior studies have generally focused on simpler problems. For example, the fused modalities are all 3D data of the same size in \cite{8515234}. In that case, multimodal features always have the same shape at each scale, so feature fusion can be easily achieved through concatenation. For ophthalmic data, the size and dimensionality of the features are modality-dependent: 3D tensors for 2D images and 4D tensors for 3D images.

\begin{figure}[!t]
\centering
\includegraphics[width=\textwidth]{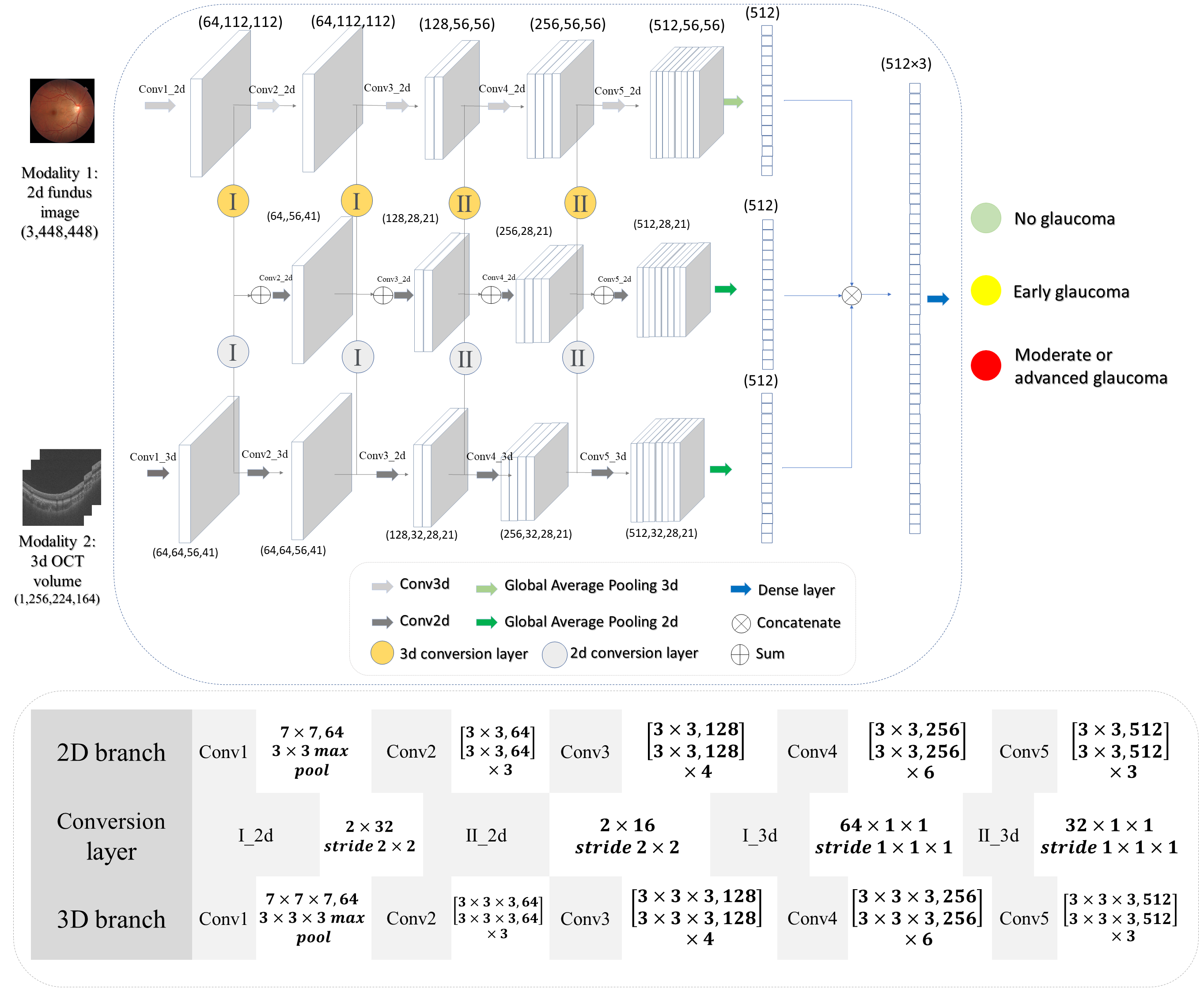}
\caption{Proposed hierarchical fusion configuration, illustrated using 2D and 3D ResNet34, for glaucoma classification from 2D fundus photography and 3D OCT. I and II are different types of conversion layers, and their configurations are shown in the list.} \label{fig1}
\end{figure}

A solution is proposed hereafter and illustrated in Fig.~\ref{fig1}. Two CNN branches are used to extract features from a multichannel 2D image and a multichannel 3D volume, respectively. Furthermore, we use a third fusion branch to achieve feature fusion at different scales. Since the dimensional features are of different dimensions and sizes, we use additional conversion convolutional layers to harmonize their shape before concatenating them. In these conversion convolution layers, the parameters are calculated according to the size of the modality-specific features.

\begin{small} 
\begin{gather*} 
F_{3D} 
\left(  C \times Z_{3D} \times X_{3D} \times Y_{3D}  \right)  \Longrightarrow    F'_{3D} \left( C \times 1 \times X_{3D} \times Y_{3D}  \right) \\
F_{2D}\left(  C \times X_{2D} \times Y_{2D}    \right)  \longrightarrow    F'_{2D} \left( C \times X_{3D} \times Y_{3D}    \right)
\end{gather*}
\end{small}

where $F$ is feature of modality, $X,Y,Z,C$ represent the length, width, depth, and number of channels of the features. $\Longrightarrow$ and $\longrightarrow$ represent the 3D and 2D conversion convolutional layers respectively. The convolution kernel size and stride of 3D conversion convolutional layers are ($Z_{3D} \times 1 \times 1$) and (1 $\times 1 \times 1$). For the 2D conversion layer, the stride is set to (2, 2) and the filter size is set to $(X_{2D} - 2 [X_{3D} - 1], Y_{2D} - 2 [Y_{3D} - 1])$, without padding, to ensure that $F'_{2D}$ matches the size of $F'_{3D}$. 
The parameters of each convolutional layer are shown in Fig.~\ref{fig1} for ResNet34.

We also extract the features from the 2D CNN block to reduce the number of parameters of the fusion branch. In the end, the high-dimensional features of the three branches are concatenated, and the classification layer is used to make the final classification.

In addition to the advantages of intermediate fusion, hierarchical fusion also considers features from different scales, enhancing the correlation between different modalities and increasing the accuracy of diagnosis. 

\begin{figure}
\includegraphics[width=\textwidth]{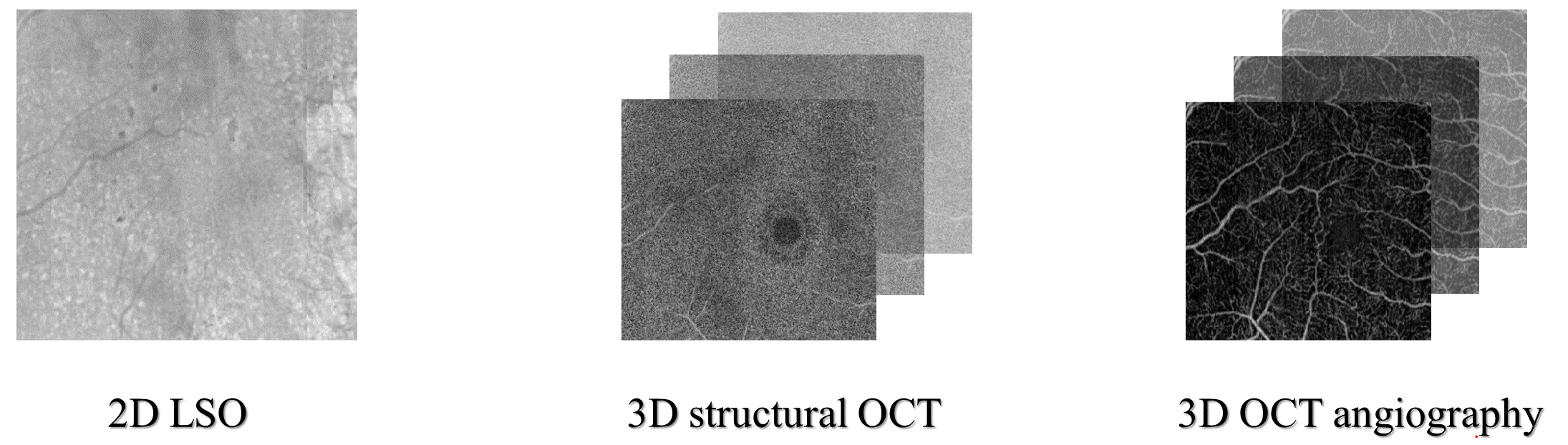}
\caption{Data from three imaging modalities in the PlexEliteDR dataset.} \label{fig2}
\end{figure}

\section{Material and experiments}

We evaluated the proposed method using the public GAMMA challenge dataset for glaucoma classification and a private dataset for proliferative DR (PDR) classification: PlexEliteDR. For the GAMMA dataset, we analyzed clinical data from 2D fundus images, and 3D OCT scans to classify glaucoma into three groups based on visual features: no glaucoma, early glaucoma, and moderate or advanced glaucoma. For the PlexEliteDR dataset, we investigated the fusion of 3 modalities: 3D structural OCT, 3D OCT angiography, and 2D line scanning ophthalmoscope (LSO) for the classification of PDR and NPDR.

\subsection{Data}

\subsubsection{GAMMA dataset}
 is provided by Sun Yat-sen Ophthalmic Center, Sun Yat-sen University, Guangzhou, China. There are 200 pairs of clinical modality images in the dataset, 100 pairs in the training set, and 100 pairs in the test set. Each pair contains a fundus image and an OCT volume. The OCT volumes were acquired using a Topcon DRI OCT Triton  machine. The OCT was centered on the macula and had a 3 $\times$ 3 mm en-face field of view. The Kowa 2000 $\times$ 2992 and Topcon TRC-NW400 cameras were used to acquire fundus images \cite{https://doi.org/10.48550/arxiv.2202.06511}.

There are 50 pairs of no glaucoma patients in the training set, 26 pairs of early glaucoma patients, and 24 pairs of moderate or advanced glaucoma patients in the training set. These pairs were divided as follows: 80 pairs for training (41 pairs no glaucoma,  21 pairs early glaucoma, 18 pairs moderate or advanced glaucoma) and 20 pairs for validation (9 pairs no glaucoma, 6 pairs early glaucoma, 5 pairs moderate or advanced glaucoma).

\subsubsection{PlexEliteDR dataset}
 is a private dataset. 3D structural OCT, 3D OCT angiography and 2D LSO data were acquired simultaneously with a Plex® Elite 9000 (Carl Zeiss Meditec Inc. Dublin, California, USA) as Fig.~\ref{fig2}.  Scanning protocols included 3 $\times$ 3 mm, 6 $\times$ 6 mm, and 15 $\times$ 9 mm. According to the International Clinical Diabetic Retinopathy Disease Severity Scale (ICDR) scale, the DR severity level was graded by a retina specialist using fundus photographs.

151 OCT volumes from 64 diabetic patients were collected for the binary classification. This collection was divided as follows: 88 acquisitions (from 31 patients) for training, 28 acquisitions (from 14 patients) for validation and 35 acquisitions (from 19 patients) for testing. Thirty acquisitions (including 16 in the train set, 5 in the validation set and 9 in the test set) had PDR .

\subsection{Data pre-processing}

The original 2D and 3D images were too large to train a fusion network. They were therefore cropped to remove black areas and resized. The following dimensions were used: $X = 224$, $Y = 164$ and $Z = 256$ for GAMMA, $X = Y = Z = 100$ for PlexEliteDR. For intermediate and hierarchical fusion, 2D images could be larger than 3D images: they were resized to 448 $\times$ 448 pixels for GAMMA and 400 $\times$ 400 pixels for PlexEliteDR. Note that 2D and 3D data are not spatially registered in GAMMA; they are only approximately centered on the same anatomical structure (the optic nerve head). All modality are natively registered in PlexEliteDR.

\subsection{Implementation details}

Experiments were performed using 2D and 3D versions of ResNet \cite{https://doi.org/10.48550/arxiv.1512.03385} and DenseNet \cite{8515234}. These networks were used as is, or adapted for each fusion strategy. To augment the data, RandomGamma, GaussianNoise, and flipping were applied for all tests. Gradient descent was performed with the Adam optimizer, which has an initial learning rate of 1e-4, and a weight decay rate of 1e-4.

\section{Results}

\subsection{GAMMA dataset}
We tested the performance of four ResNet networks on the same dataset: ResNet34, ResNet50, ResNet101, and ResNet152. As a standard evaluation metric for the multi-category classification task, Cohen's Kappa was used to evaluate the GAMMA dataset's three-category results.

The best-performing models were selected from the validation set and tested on the 100 pairs test set. The final Kappa results on the test set were computed independently by the PaddlePaddle deep learning platform\footnote[1]{\url{ https://aistudio.baidu.com/aistudio/competition/detail/119/0/introduction}}, which is the host platform for the GAMMA challenge.

We tested each modality separately, as well as the three fusion methods, and the results are shown in Table~\ref{tab1}.

\begin{table}
\centering
\caption{Kappa results of different fusion methods on the GAMMA dataset}\label{tab1}
\begin{tabular}{cccccc}
\toprule  
Backbone & \makecell[c]{Single modality \\ (fundus image)} & \makecell[c]{Single modality \\ (OCT)}
 & Early fusion & \makecell[c]{Intermediate\\ fusion}  & \makecell[c]{Hierarchical\\ fusion} \\
\midrule  
ResNet34 &  0.6997  &0.6841 &0.6718  & 0.7547 & 0.7684\\
ResNet50 &  0.6555 & 0.5952& 0.6896& 0.7690 & \textbf{0.8404}\\
ResNet101 & 0.6767  & 0.5794& 0.7113& 0.7551 & 0.8255\\
ResNet152 &  0.5207 & 0.4646&0.4642 & 0.6570 & 0.7816\\
\midrule 
Average &  0.6382 & 0.5808 &0.6342 & 0.7340 & 0.8040\\
\bottomrule 
\end{tabular}
\end{table}

The Kappa results above show that color fundus images outperform OCT volumes when using data from a single modality. In addition, ResNet34 has better performance, possibly because simple features of a single modality are easy to learn. Although, according to the average of different backbones, 0.6382 is still far from a result that can be useful for diagnosis. Thus, single-modality glaucoma classification is very ineffective.

Results for the early fusion were not significantly improved. The reason probably is that fundus and OCT images are not spatially registered in this dataset.

Intermediate fusion is a more suitable fusion algorithm in this case because of the disparity between fundus images and OCT volumes, and the dual feature extraction branch can effectively handle the large differences between modalities. As a result, the performance of intermediate fusion is greatly improved compared to the single-modality scenario. In addition, for ResNet152, we had to reduce the batch size during training to avoid the device from exceeding the memory limit, which is one reason for the poor performance of ResNet152.

The GAMMA challenge also uses intermediate fusion as its baseline \cite{https://doi.org/10.48550/arxiv.2202.06511}. In the official baseline, two convolutional branches are used for intermediate fusion. Based on 3D OCT, retinal thickness is used as a channel for the input of the 2D convolutional branch in the algorithm. By contrast, we utilize 3D convolutional branches to extract 3D OCT features, which allows us to fully utilize the spatial features of 3D data. This is why our Kappa value of 0.734 for intermediate fusion is higher than the official intermediate fusion result of 0.702.

 Comparatively to intermediate fusion, hierarchical fusion is able to better exploit correlations between features of different dimensions: the Kappa value increased by 0.0700. These results support the efficiency of our hierarchical fusion.

Specifically, our hierarchical fusion performs very well on ResNet50 and ResNet101. To achieve a higher score in the GAMMA challenge, we selected the models of ResNet50 and ResNet101 for further training. The training and validation sets were re-divided and the checkpoint obtained from the previous test was fine-tuned. Finally, we achieved a Kappa value of 0.8662 for ResNet50 and 0.8745 for ResNet101. For our hierarchical fusion, we improved the final Kappa to 0.8996 by ensembling the predicted values of ResNet50 and ResNet101 models.

\subsection{PlexEliteDR dataset}

For the PlexEliteDR dataset, the following backbones were investigated for each method: ResNet50, ResNet101, DenseNet121, and DenseNet169. The Area under the ROC Curve (AUC) was used to assess the binary classification performance.

\begin{table}
\centering
\caption{Results of different fusion methods on the PlexEliteDR dataset}\label{tab3}
\begin{tabular}{cccccc}
\toprule  
Method& Backbone& AUC & Sensitivity & Specificity& Improvement\\
\midrule  
Single modality (Structure)&  ResNet101 &0.859 & 0.78 & 0.77& Baseline \\
Single modality (Flow)&  DenseNet169 &0.816 & 0.78 & 0.85&-0.043\\
Single modality (LSO)&  DenseNet121 &0.662 & 0.67 & 0.74&-0.197\\
Hierarchical fusion &  DenseNet121 &\textbf{0.911} & \textbf{0.86} & \textbf{0.88}&\textbf{+0.052}\\
Early Fusion &  DenseNet121 &0.865 & 0.78 & 0.85&+0.006\\
Intermediate Fusion & DenseNet121 &0.744 & 0.67 & 0.85&-0.115\\
\bottomrule 
\end{tabular}
\end{table}

Using a single modality, the structure data achieved the best performance: AUC reaches 0.859 using ResNet101 (this is our baseline). Intermediate fusion performed worse than baseline. Unlike the GAMMA dataset, the three modalities are spatially aligned in PlexEliteDR, so the early fusion approach achieves good results. Hierarchical fusion achieves the best results: AUC reaches 0.911 using DenseNet121. The LSO images do not provide very distinct pathological details, compared to fundus images, hence a more limited impact of information fusion.

\section{Conclusion}
This paper presents three fusion strategies based on deep learning: early fusion, intermediate fusion, and hierarchical fusion. On glaucoma and diabetic retinopathy classification tasks, they clearly outperform classification using a single modality. The novel hierarchical fusion approach is particularly promising, both for glaucoma grading and proliferative DR detection. However, these experiments should be replicated in larger datasets to demonstrate clinically useful detection performance. Additionally, hierarchical fusion is a complex model, and the larger number of parameters requires a robust hardware setup. Larger input sizes are worth testing as hardware evolves.

\section*{Acknowledgements}
The work takes place in the framework of the ANR RHU project Evired. This work benefits from State aid managed by the French National Research Agency under ``Investissement d'Avenir'' program bearing the reference ANR-18-RHUS-0008.

%
%
%

\bibliographystyle{splncs04}
\bibliography{paper8}

\end{document}